\documentclass[reprint,aps,nofootinbib,preprintnumbers,amsmath,amssymb,11pt]{revtex4}

\usepackage{amsmath}
\usepackage{amssymb}
\usepackage{setspace}
\usepackage{amsfonts}
\usepackage{graphicx}
\usepackage{dcolumn}
\usepackage{bm}
\usepackage{natbib}
\newcommand{\half}{\frac{1}{2}}
\newcommand{\beq}{\begin{equation}}
\newcommand{\eq}{\end{equation}}
\newcommand{\bea}{\begin{eqnarray}}
\newcommand{\ea}{\end{eqnarray}}
\newcommand{\p}{\partial}
\newcommand{\nn}{\nonumber}
\begin{document}
\preprint{MIFPA-10-14}
\preprint{DAMTP-2010-24}

\title{\LARGE  Extremal Three-point Correlators in Kerr/CFT\\}

\author{\vspace*{1 cm}\large
Melanie Becker$^\dagger$, Sera Cremonini$^{\dagger,\S}$, Waldemar Schulgin$^\dagger$}
\email{mbecker,sera,schulgin@physics.tamu.edu}
\affiliation{\vspace*{0.5 cm}$^\dagger$ George and Cynthia Mitchell Institute for Fundamental Physics and Astronomy,
Texas A\&M University, College Station, TX 77843--4242, USA\\}
\affiliation{$^\S$ Centre for Theoretical Cosmology, DAMTP, CMS,\\
University of Cambridge, Wilberforce Road, Cambridge, CB3 0WA, UK}

\begin{abstract}
\vspace*{1 cm}
We compute three-point correlation functions in the near-extremal, near-horizon region of a Kerr black hole,
and compare to the corresponding finite-temperature conformal field theory correlators.
For simplicity, we focus on scalar fields dual to operators ${\cal O}_h$ whose conformal dimensions
obey $h_3=h_1+h_2$,
which we name \emph{extremal} in analogy with the classic $AdS_5 \times S^5$ three-point function in the literature.
For such extremal correlators we find perfect agreement with the conformal field theory side,
provided that the coupling of the cubic interaction contains a vanishing prefactor $\propto h_3-h_1-h_2$.
In fact, the bulk three-point function integral for such extremal correlators diverges as $1/(h_3-h_1-h_2)$.
This behavior is analogous to what was found in the context of
extremal AdS/CFT three-point correlators.
As in the AdS/CFT case our correlation function can nevertheless be computed via analytic continuation
from the non-extremal case.
\end{abstract}

\vspace*{3 cm}
\maketitle

\newpage
\tableofcontents

\def\thesection{\arabic{section}}
\def\thesubsection{\arabic{section}.\arabic{subsection}}
\numberwithin{equation}{section}

\section{Introduction}

Since its inception, the gauge/gravity duality has proven to be a rich avenue for gaining insight
into quantum gravity -- leading, for instance, to a microscopic understanding of black hole entropy.
It has also given us a new perspective on the strong coupling dynamics of many gauge theories.
While it was originally best understood in the context of ${\cal N}=4$ SYM,
it has now been applied to a wide spectrum of strongly coupled systems,
ranging from the quark gluon plasma to condensed matter physics.
Even more recently, the gauge/gravity correspondence has turned out to play a role in the description of
extreme Kerr black holes -- the near horizon region of such geometries has been shown to have a dual
description in terms of a chiral two-dimensional conformal field theory
This Kerr/CFT correspondence \cite{Guica:2008mu} is particularly interesting given that many
astrophysical black holes
increase their spin while accreting matter, and tend to approach the extreme Kerr limit.

The boundary theory dual to the near-horizon region of an \emph{extreme} Kerr black hole
is the left-moving sector of a two-dimensional conformal field theory, with central charge
$c_L = 12 J/\hbar$.
Agreement between Cardy's entropy formula on the CFT side and
the gravity-side Bekenstein-Hawking entropy has provided evidence for the correspondence \cite{Guica:2008mu}.
The analysis has been successfully extended in a number of ways and applied to a variety of
extremal rotating black holes
\cite{Lu:2008jk,Hartman:2008pb,Azeyanagi:2008dk,Compere:2009dp,Krishnan:2009tj,
Balasubramanian:2009bg,Castro:2009jf,Cvetic:2009jn,Jejjala:2009if,Chow:2008dp,
Isono:2008kx,Azeyanagi:2008kb,Chen:2009xja,Hotta:2009bm, Ghezelbash:2009gy,Garousi:2009zx,
Azeyanagi:2009wf,Wu:2009di,Chen:2009ht,Peng:2009ty,Loran:2009cr,Ghezelbash:2009gf,Lu:2009gj,Matsuo:2009sj,Peng:2009wx,Mei:2010wm}.
However, extending Kerr/CFT to the \emph{near-extreme} Kerr black hole, where right-moving
excitations are also turned on, has proven challenging.
The main obstacle comes from the fact that consistent boundary conditions which allow for
both left- and right-movers have not been found yet (see \cite{Bredberg:2009pv}).
Recently, \cite{Bredberg:2009pv} has followed an alternate route for testing Kerr/CFT in the near-extreme case,
matching the (near-extreme) black hole absorption cross section to the two-point correlation function
of a 2D non-chiral conformal field theory. We wish to extend these studies to a more systematic analysis by calculating
three-point functions, with the goal of testing the correspondence.

In the early days of AdS/CFT, studies of correlation functions proved to be a valuable tool for testing the
conjecture, but were done at zero temperature and relied on having a certain amount of supersymmetry.
The precise dictionary between bulk and boundary physics was developed mainly for Euclidean signature --
a Lorentzian prescription \cite{Son:2002sd,Herzog:2002pc,Iqbal:2009fd}, appropriate for real-time, finite temperature correlators was only put forth years
later, and it still hasn't been fully exploited for general n-point correlation functions.
While the Lorentzian type prescription for computing two-point functions is now widely used \cite{Son:2002sd,Iqbal:2009fd},
and a formalism generalizing \cite{Herzog:2002pc} has been proposed for dealing with
higher correlation functions \cite{Skenderis:2008dg}, there is very little work -- if any -- on
finite-temperature three-point correlation functions (or higher) in the context of AdS/CFT.

In this paper, we begin to address this issue, by computing three-point correlators in the
near-extremal, near-horizon region of a Kerr black hole, the so-called near-NHEK geometry.
While part of our motivation is to better understand how to deal with finite-temperature three-point
correlation functions in the gauge/gravity duality, our main interest is in testing the Kerr/CFT correspondence
in the presence of non-chiral excitations.
We consider three-point correlators of a scalar field $\Phi_h$ in the near-NHEK geometry,
assumed to be dual to a CFT operator ${\cal O}_h$ of conformal weight $h$,
with a cubic interaction of the form
\beq
\mathcal{G} \int \Phi_{h_1} \Phi_{h_2} \Phi_{h_3} \, .
\eq
In this article we focus for simplicity on the special case where the conformal weights are constrained to be
$h_3=h_1+h_2$, and leave the treatment of arbitrary weights to future work \cite{becker}.
We refer to this case as the ``extremal'' limit, in analogy with classic $AdS_5 \times S^5$ computations \cite{D'Hoker:1999ea}, where
the three-point function vanished unless $h_i +h_j \geq h_k$ for any pair of weights,
and the extremal correlator saturated this bound.
In the extremal case the three-point function reduces to the product of two two-point functions,
hence its simplicity.
While this is clearly an overly simplified correlator, it will still give rise to interesting features.
In particular, we find perfect agreement with the CFT finite-temperature three-point correlator.
However, as we will see, our computation seems to require a very specific form of the coupling of the
cubic interaction. We discuss this in detail in Section 4.

The outline of the paper is as follows.
We start in Section 2 with a review of the four-dimensional Kerr geometry, focusing on its near-horizon
region, both in the extremal and near-extremal regimes. We also briefly outline the greybody factor
computation of \cite{Bredberg:2009pv}, and its agreement with the CFT Green's function.
In Section 3 we outline the standard zero-temperature AdS/CFT prescription for computing correlators, as well
as the subtleties of the finite temperature case. After introducing a proposal for the
bulk-to-boundary propagator
in near-NHEK, we show that it reproduces the two-point function calculation of \cite{Bredberg:2009pv}.
Section 4 is devoted to the main results of our paper, the computation of the three-point correlators,
on the gravity side as well as on the finite temperature CFT side.
We conclude in Section 5 with a discussion of our results.

\section{Kerr/CFT}
\label{KerrCFTSection}

\subsection{Kerr, NHEK and near-NHEK}

Our main interest in this paper is in computing three-point functions for the so-called
\emph{near-NHEK} geometry, the near-horizon region of a near-extremal Kerr black hole.
Before moving on to the correlation function computations, we review briefly some of the basic
properties of the four-dimensional Kerr solution, focusing in particular on how one can obtain the
near-horizon geometry, both in the extremal and near-extremal case.
We will also remind the reader of the results of \cite{Bredberg:2009pv}, where the finite-temperature CFT
two-point function was shown to match the gravitational computation of greybody factors in the NHEK
geometry.
In this section we will follow \cite{Bredberg:2009pv} very closely.

The 4D Kerr black hole, which is parametrized by the ADM mass $M$ and angular momentum $J$, is described by
\beq
\label{Kerr}
ds^{2}_{\;\text{Kerr}}= -\frac{\Delta}{R^2}\left(d\hat{t}-a\sin^2\theta d\hat{\phi} \right)^2 +
\frac{\sin^2\theta}{R^2} \left(\left(\hat{r}^2+a^2\right)d\hat{\phi}-a d\hat{t} \right)^2 +
\frac{R^2}{\Delta} d\hat{r}^2 +R^2 d\theta^2 \; ,
\eq
where
\beq
a= \frac{J}{M} \, , \quad \quad R^2 = \hat{r}^2+a^2 \cos^2\theta
\quad \quad \mbox{and} \quad \quad \Delta = \hat{r}^2 - 2 M \hat{r} +a^2 \, ,
\eq
in units of $G_4=\hbar=c=1$.
The inner and outer horizons are related to the mass $M$ via
\beq
r_{\pm}=M \pm \sqrt{M^2-a^2} \, .
\eq
Away from extremality the Hawking temperature, the horizon angular velocity and the Bekenstein-Hawking
entropy are given by
\beq
T_H = \frac{1}{8\pi M} \frac{r_+-r_-}{r_+} \, , \quad \quad \Omega_H = \frac{a}{2Mr_+}\, , \quad \quad
S=2\pi M r_+ \, .
\eq
The extremal limit, in which the two horizons coincide and the Hawking temperature vanishes,
corresponds to $a=M$ and thus $r_+=M$. In that case the black hole carries the maximum amount
of angular momentum\footnote{The solution has naked singularities unless $-M^2 \leq J \leq M^2$.},
$J=M^2$, and the entropy takes the simple form
\beq
\label{SBH}
S_{\text{ext}}=2\pi J \, .
\eq

Since the proper spatial distance to the horizon of the extreme Kerr geometry is infinite, one can
zoom into its near-horizon region.
More precisely, the near-horizon region can be isolated by
choosing new coordinates
\beq
\label{NHEKcoords}
t = \lambda \, \frac{\hat{t}}{2M} \, , \quad \quad r=\frac{1}{\lambda} \, \frac{\hat{r}-M}{M} \, , \quad \quad
\phi = \hat{\phi}- \frac{\hat{t}}{2M} \, ,
\eq
and taking the $\lambda \rightarrow 0$ limit \cite{Bardeen:1999px,Hartman:2008pb}.
The resulting near-horizon extreme Kerr (NHEK) geometry is described by
\beq
\label{NHEK}
ds^2_{\;\text{NHEK}}= 2 J \,\Gamma(\theta) \Bigl(-r^2 dt^2 + \frac{dr^2}{r^2} + d\theta^2 +
\Lambda(\theta)^2 (d\phi + r dt)^2 \Bigr)\, ,
\eq
where
\beq
\label{AngDep}
\Gamma(\theta)=\frac{1+\cos^2\theta}{2} \quad \quad \text{and} \quad \quad
\Lambda(\theta)= \frac{2\sin\theta}{1+\cos^2\theta} \, .
\eq
Since in the Kerr/CFT correspondence the dependence on the angle $\theta$ is \emph{frozen},
the explicit form of $\Gamma(\theta)$ and $\Lambda(\theta)$ will not play a role in the rest of the discussion.

NHEK describes a $U(1)$ fibration over $AdS_2$.
As can be seen from (\ref{NHEKcoords}), the boundary of this geometry is at $r=\infty$, which corresponds
to the entrance of the throat where the NHEK region glues onto the full asymptotically flat geometry.
Furthermore, NHEK has an $SL(2,\mathbb{R})_R \times U(1)_L$ isometry group.
In fact, it was shown in \cite{Guica:2008mu} that with consistent boundary conditions the $SL(2,\mathbb{R})_R$
becomes trivial, while the $U(1)_L$ is enhanced to a Virasoro algebra with central charge $c_L=12J$.
Thus, it is precisely the enhancement of the $U(1)_L$ that is responsible for the presence of
the dual 2D chiral CFT.
The quantum theory in the Frolov-Thorne vacuum has a left-moving temperature\footnote{This temperature can also be read off from the first law of thermodynamics.}
\beq
T_L = \frac{1}{2\pi} \, ,
\eq
which, combined with the expression for the central charge $c_L$ and Cardy's entropy formula,
yields the entropy of the dual conformal field theory:
\beq
S_{\text{CFT}} = \frac{\pi^2}{3} c_L T_L = 2\pi J \, .
\eq
This matches the Bekenstein-Hawking entropy (\ref{SBH}) of the extreme Kerr solution.

The near-horizon limit of a Kerr black hole can be further modified in such a way to maintain some
excitation energy above extremality \cite{Bredberg:2009pv}.
This can be done by taking an appropriate limit of the Kerr geometry, where
$T_H \rightarrow 0$ and $\hat{r} \rightarrow r_+$ while the dimensionless
\emph{near-horizon} temperature
\beq
T_R \equiv \frac{2MT_H}{\lambda}=\frac{\tau_H}{4\pi \lambda} \,
\eq
is held fixed as $\lambda \rightarrow 0$.
Physically, this means that while the Hawking temperature at asymptotic infinity is zero,
the temperature measured near the horizon remains finite.
The coordinate change needed to obtain the near-extremal metric \cite{Bredberg:2009pv} is a modification of
(\ref{NHEKcoords}):
\beq
t = \lambda \, \frac{\hat{t}}{2M} \, , \quad \quad r=\frac{1}{\lambda} \, \frac{\hat{r}-r_+}{r_+} \, ,
\quad \quad \phi = \hat{\phi}- \frac{\hat{t}}{2M}\, .
\eq
The resulting near-NHEK geometry is described by the metric:
\beq
\label{nearNHEK}
ds^2_{\;\text{near-NHEK}}= 2 J \,\Gamma \Bigl(-r(r+4\pi T_R) dt^2 + \frac{dr^2}{r(r+4\pi T_R)} + d\theta^2 +
\Lambda^2 (d\phi + (r+2\pi T_R) dt)^2 \Bigr)\, ,
\eq
with $\Gamma(\theta)$ and $\Lambda(\theta)$ still given by (\ref{AngDep}).

The near-NHEK geometry is globally diffeomorphic to NHEK.
However, as explained in \cite{Bredberg:2009pv}, the coordinate transformation which relates
near-NHEK to NHEK, and eliminates dependence on
the near-horizon temperature $T_R$, is singular at the boundary --
the boundary regions where the far region is glued to the near region are not
diffeomorphic for NHEK and near-NHEK.
Obtaining the near-NHEK geometry requires a delicate limiting procedure
(see \cite{Castro:2010vi} for a recent discussion), which might be analogous to having a cutoff
on the radial coordinate.
Thus, we will think of the possible CFT dual to near-NHEK as living on some
``effective'' boundary which is rather close to the black hole horizon, which we
denote by $r_{\text{cutoff}}$.

\subsection{Macroscopic and Microscopic Greybody Factors}

As pointed out in \cite{Bredberg:2009pv}, only certain modes survive the $T_H \rightarrow 0$ limit required
to obtain the near-NHEK geometry -- modes with a frequency close to the superradiant bound
\footnote{To see why this is so, one can look at the expression for the
Boltzmann factor which enters the Hawking decay rate.
If a scalar in the original Kerr geometry has a behavior of the form $\sim e^{-i\hat{\omega}\hat{t} +i m\hat{\phi}}$,
in the small Hawking temperature limit
the Boltzmann factor
$\sim e^{-\frac{\hat{\omega}-m\Omega_H}{T_H}}$
will suppress all modes except
for ones with $\hat{\omega}-m\Omega_H \sim 0$, \emph{i.e.} superradiant modes.}.
The study of superradiant scattering off a Kerr black hole has a long history \cite{Starobinsky:1973,StarobinskyAndChurilov:1973,Teukolsky:1973ha,Press:1973zz,Teukolsky:1974yv,Futterman:1988ni,zeldovich}.
Scattering of a scalar field in the near-NHEK geometry has been revisited in \cite{Bredberg:2009pv,Cvetic:2009jn}
, where it is shown that the black hole absorption cross section $\sigma_{\text{abs}}$
-- hence the greybody factor --
is reproduced by the finite-temperature two-point correlation function of a
\emph{non-chiral} 2D conformal field theory.
For a detailed, self-contained derivation of $\sigma_{\text{abs}}$ and its CFT counterpart we refer the reader to
\cite{Bredberg:2009pv}. Here we extract only the ingredients which we will need throughout
our analysis.

Taking the scalar field incident on the black hole to be of the form
\footnote{Note that our $\omega$ is $n_R$ in the notation of \cite{Bredberg:2009pv}.}
\beq
\label{scalar}
\Phi(t,r,\phi,\theta) = e^{-i \omega t + i m \phi} \psi(r)S(\theta) \, ,
\eq
and solving the near-NHEK wave equation,
the radial wavefunction which is \emph{ingoing} at the horizon can be shown to be
\beq
\label{fullsol}
\psi^{in}_{m,\omega}(\tilde{r}) = N \tilde{r}^{-\frac{i}{2}(m+\frac{2 \omega}{\tau_H})}
\Bigl(1+\frac{\tilde{r}}{\tau_H}\Bigr)^{-\frac{i}{2}(m-\frac{2 \omega}{\tau_H})}
\, F\left(\half+\beta-im,\half-\beta-im,1-i\left(m+\frac{2 \omega}{\tau_H}\right),-\frac{\tilde{r}}{\tau_H}\right)\, ,
\eq
where $\tilde{r} \equiv \lambda \,r$,
$N$ is the normalization factor and
$\beta^2=K_l-2m^2+1/4$ \footnote{Here $K_l$ is
the separation constant used to separate the $r$ and $\theta$ equations of motion.}.

A standard scattering calculation then shows that the absorption cross section for propagation
in the near-horizon region takes the form
\beq
\label{abs}
\sigma_{\text{abs}} = \text{Im}(G_R) \, , \quad \text{with} \quad
G_R \sim \tau_H^{2\beta} \,\frac{\Gamma(-2\beta)\, \Gamma(\half+\beta-im)\,\Gamma(\half+\beta-i\frac{2\omega}{\tau_H})}
{\Gamma(2\beta)\,\Gamma(\half-\beta-im)\,\Gamma(\half-\beta-i\frac{2\omega}{\tau_H})} \, ,
\eq
where $G_R$ is the retarded propagator.

Matching with the conformal field theory side can be checked by computing the contribution of
the right- and left-movers to the Euclidean Green's function.
Each sector should contribute
\beq
\label{EuclG}
G_E(\omega_E) =
C_h \int_0^{1/T} e^{i\omega_E} \tau \Bigl[\frac{\pi T}{\sin \pi T \tau}\Bigr]^{2h} d\tau
\sim C_h
\frac{T^{2h-1}\, e^{i\omega_E/2T}\,\Gamma(1-2h)}{\Gamma(1-h+\frac{\omega_E}{2\pi T})\,\Gamma(1-h-\frac{\omega_E}{2\pi T})},
\eq
to the total Green's function $G_E(\omega_R,\omega_L)$,
where $\tau$ and $\omega_E$ are, respectively, the Euclidean time and frequency, while $C_h$ is a constant that
depends on the conformal dimension.
The Euclidean  and retarded Green's functions are expected to agree
at the discrete (Matsubara) frequencies
$\omega_E^{(k)}=2\pi k \,T$, $k \,\epsilon \,\mathbb{Z}$,
at which $G_E(\omega_E)$ is defined, \emph{i.e.}
\beq
\label{continuation}
G_R(i\omega_R,i\omega_L)=G_E(\omega_R,\omega_L) \, .
\eq
This agreement was in fact checked by \cite{Bredberg:2009pv}, and subsequently in \cite{Chen:2010ni}, thus showing that the macroscopic greybody factor
(\ref{abs}) is reproduced by the Green's function of a finite temperature non-chiral 2D conformal field theory,
with both right- and left-movers turned on.
In order for the matching to work, the parameters of the two theories need to be identified in the
following way:
\beq
h_L=h_R=\frac{1}{2} + \beta \, , \quad T_L=\frac{1}{2\pi}\, , \quad T_R=T_R \, ,
\quad \omega_L=m \, , \quad \omega_R=\omega\, ,
\eq
where $h$ is the conformal dimension of the CFT operators.

\section{AdS/CFT prescription for correlators}
\subsection{Prescriptions}

Most work on correlation function calculations in the AdS/CFT literature was done
using an Euclidean formulation of the conjecture (see e.g. \cite{Aharony:1999ti} for a nice review).
At the core of the AdS/CFT correspondence in Euclidean space is the equivalence of partition functions
\beq
\label{adscft}
Z_{CFT}\,[\phi_0(\vec{x})]=\langle e^{-\int \phi_0(\vec{x}) \;{\cal O}(\vec{x})\;}\rangle_{CFT} =
Z_{string}[\phi_0(\vec{x})] \sim e^{-S_{sugra}[\phi_0(\vec{x})]} \, ,
\eq
which states that the classical supergravity action serves as a generating functional for correlation functions
of gauge-invariant operators ${\cal O}$ in the boundary field theory.
The role of the source is played, on the gravity side, by the boundary value $\phi_0(\vec{x})$ of the bulk field $\phi$
that couples to ${\cal O}$.
One then computes correlators in the usual field theory sense by taking
functional derivatives $\frac{\delta}{\delta \phi_0} \, e^{-S_{sugra}[\phi_0]}$.

A key role in the AdS/CFT correspondence is played by the so-called bulk-to-boundary propagator
$K(r,\vec{x};\vec{x}^{\, \prime} )$, which explicitly connects the bulk scalar field to its boundary value:
\beq
\phi_h(r,\vec{x})=\int_{\p M} d\vec{x}^{\, \prime}  K_h(r,\vec{x};\vec{x}^{\, \prime} )\; \phi_0(\vec{x})\, .
\eq
We can read this as encoding information on how a bulk field is generated in response
to the boundary source $J \equiv \phi_0(\vec{x})$.
In addition to being a solution to the wave equation $\Box \, K_h=0$,
the bulk-to-boundary propagator $K_h$
must also respect the way in which the scalar field behaves at the boundary.
More precisely, it must obey
\beq
\label{Kscaling}
r^{d-h} K_h(r,\vec{x};\vec{x}^{\, \prime} ) \rightarrow \delta^d(\vec{x}-\vec{x}^{\, \prime} ),
\eq
if the scalar field scales as
\beq
\phi(r,\vec{x})\rightarrow r^{h-d}\phi_0(\vec{x}),
\label{scaling}
\eq
as it approaches the boundary.

Precisely because the prescription (\ref{adscft}) involves functional derivatives with respect to the
boundary value of the field, $\frac{\delta S_{Sugra}}{\delta \phi_0}$, the bulk-to-boundary propagator
is a crucial ingredient in the computation of correlators.
For instance, the $AdS$ two-point function calculation reduces
to the evaluation of a boundary term of the schematic form
\beq
\label{Eucl2pt}
\langle {\cal O}(x) {\cal O}(y) \rangle \sim
\frac{\delta^2 S_{sugra}}{\delta \phi_0(x) \; \delta \phi_0(y)}
\sim \int_{\p M} d\vec{x}^{\, \prime} \,
K_h(r,\vec{x}^{\, \prime} ;\vec{x})  \, \frac{\p}{\p r} K_h(r,\vec{x}^{\, \prime} ;\vec{y})
\, \sim \, \frac{c_h}{|\vec{x}-\vec{y}|^{2h}},
\eq
where
the constant $c_h$ encodes the dependence on $h$.
Expressions for n-point correlation functions can be generated in an analogous way.
In particular, the 3-point function is given by a product of three bulk-to-boundary propagators, each leg
connecting the same point in the bulk to a point on the boundary where the CFT operator is inserted:
\beq
\label{3ptprescription}
\langle O_{h_1}(\vec{x}_1)\, O_{h_2}(\vec{x}_2)\, O_{h_3}(\vec{x}_3)\rangle
\sim \int d\vec{x}  \, dr \,  K_{h_1}(r,\vec{x} ;\vec{x}_1)
K_{h_2}(r,\vec{x} ;\vec{x}_2) K_{h_3}(r,\vec{x} ;\vec{x}_3) \, .
\eq
Notice that this is no longer just a boundary term, as in the 2-point function case -- here one integrates
over the entire bulk.

The previous prescription is the one mostly used in the literature in order to calculate zero temperature
correlation functions in the context of AdS/CFT.
While the prescription (\ref{adscft}) works well for Euclidean correlators, a number of subtleties arise in
Lorentzian signature, which need to be taken into account.
In the Euclidean case one finds only non-normalizable modes, and the classical solution $\phi$ is
uniquely determined by the requirement of
regularity in the interior and by its value $\phi_0$ at the boundary.
On the other hand in Lorentzian space -- where there are normalizable as well as non-normalizable modes --
requiring regularity at the horizon is insufficient, and one needs to impose a more refined boundary condition.
The correct choice of normalizable modes is crucial, since it specifies the vacuum of the AdS space \cite{Balasubramanian:1998sn,Balasubramanian:1998de,Giddings:1999qu,Satoh:2002bc,Marolf:2004fy}.
This is known to be a reflection of the multitude of real time Green's functions (Feynman, retarded, advanced)
in finite temperature field theory. These issues were discussed -- and resolved in the context of
two-point correlators -- in \cite{Son:2002sd,Herzog:2002pc}.
In particular, \cite{Son:2002sd} put forth a prescription for computing Minkowski two-point Green's functions
from gravity, which has become widely utilized (see also the prescription of \cite{Iqbal:2009fd}).

Let's outline the Minkowski two-point function recipe \cite{Son:2002sd} for the case of a
scalar field $\phi$ in a black hole background.
The solution to the wave equation which satisfies incoming boundary conditions
at the black hole horizon
\footnote{For zero temperature AdS, the second boundary condition is the requirement of regularity at the interior
of AdS. For the finite temperature case, it is an incoming (or outgoing) wave boundary condition, corresponding,
respectively, to the retarded (or advanced) CFT correlator.}
can be written
\emph{near the boundary} as
\beq
\label{phibd}
\phi(r,\vec{x})={\cal A} \, r^{h-d}(1+\ldots) + {\cal B} \, r^{-h}(1+\ldots)\; ,
\eq
where the exponents are related to the conformal dimension $h$ of the dual operator ${\cal O}$,
and $d$ is the dimension of the boundary theory.
More specifically ${\cal A}$, which represents the contribution from the non-normalizable mode,
is interpreted as a source for the operator ${\cal O}_h$.
The recipe for Minkowski correlators then gives, for the retarded two point function of the operator ${\cal O}_h$,
\beq
\label{GRet}
G_R \sim \frac{{\cal B}}{{\cal A}} + \mbox{contact terms} \, .
\eq
We emphasize that to compute $G_R$ all that was needed was the asymptotic behavior of the wavefunction.
The Minkowski recipe \cite{Son:2002sd} for the two-point function entails only a slight modification of the
Euclidean prescription (\ref{Eucl2pt}),
where one must take into account appropriate (incoming) boundary conditions, and evaluate the flux
$\sim K \p K$ only at the asymptotic boundary.
In the next section we will show in detail how this prescription is implemented, for the case of Kerr/CFT.

Very surprisingly, to our knowledge no explicit calculation of finite temperature n-point correlation functions
(besides $n=2$) has been worked out in the context of AdS/CFT.
For a generalization of the Schwinger-Keldysh formalism from two-point functions
\cite{Son:2002sd,Herzog:2002pc} to n-point correlation functions we refer the reader to \cite{Skenderis:2008dg} (see also \cite{vanRees:2009rw}).
Given our goal to calculate three-point functions in Kerr/CFT, we would like to propose a simple \emph{working
prescription} for finite-temperature n-point functions:
we claim that once the appropriate (incoming) boundary conditions are taken into account when constructing
the bulk-to-boundary propagator, a straightforward application of the Euclidean prescription (3.6)
for three point correlation functions will give the right finite-temperature conformal field theory correlator.
We emphasize that we are restricting ourselves to a comparison only for Matsubara frequencies,
for which we know that (\ref{continuation}) holds, \emph{i.e.} the retarded and Euclidean Green's functions
agree.
We will present evidence of our claim in Section 4, where the three point function is calculated in detail.

\subsection{The bulk-to-boundary propagator and the 2-point function}

As we have seen, the bulk-to-boundary propagator $K_h$ is a key ingredient for the
calculation of correlation functions.
Since our proposal is to use the prescription (\ref{3ptprescription}), we first need to find the bulk-boundary propagator which is appropriate for
the near-NHEK geometry.
If the propagator is chosen correctly, in addition to satisfying (\ref{Kscaling})
it should reproduce the two-point correlation function calculation
of \cite{Bredberg:2009pv}.

We propose a bulk-to-boundary propagator of the form
\beq
\label{bulktoboundary}
K(r,t^\prime,\phi^\prime;t, \phi) = \int dm \int d \omega \; \psi(r,m,\omega) \; e^{-im(\phi-\phi^\prime)}
e^{i \omega (t-t^\prime)},
\eq
where $\psi(r,m,\omega)$ is the incoming radial wavefunction
\footnote{Notice that this is the same wavefunction of (\ref{fullsol}), except that to
simplify the notation, we have replaced $\tilde{r}$ by $r$, which we'll stick to in the rest of the paper.},
\begin{equation}
\psi^{in}_{m,\omega}(r) = N r^{-\frac{i}{2}(m+\frac{2 \omega}{\tau_H})}
\Bigl(1+\frac{r}{\tau_H}\Bigr)^{-\frac{i}{2}(m-\frac{2 \omega}{\tau_H})}
\, F\left(\half+\beta-im,\half-\beta-im,1-i\biggl(m+\frac{2 \omega}{\tau_H}\biggr),-\frac{r}{\tau_H}\right)
\, .
\end{equation}
The wavefunction has an asymptotic expansion of the form (\ref{phibd}),
\bea
\psi^{in}_{m,\omega} &\sim& N \Bigl[ {\cal A} \Bigl(r^{-\half+\beta} + {\cal O}(r^{-3/2+\beta})\Bigr) +
{\cal B} \Bigl( r^{-\half-\beta} + {\cal O}(r^{-3/2-\beta})\Bigr) \Bigr]\, ,
\ea
with the normalization and the remaining constants given by:
\bea
\label{constants}
N &=& \frac{1}{{\cal A}} \, , \nn \\
{\cal A} &=& \frac{\Gamma(2\beta)\, \Gamma(1-im-i\frac{2\omega}{\tau_H})}
{\Gamma(\half+\beta-im)\,\Gamma(\half+\beta-i\frac{2\omega}{\tau_H})} \,
\tau_H^{\half-\beta-\frac{i}{2}(m+\frac{2\omega}{\tau_H})} \, , \nn \\
{\cal B} &=& \frac{\Gamma(-2\beta)\, \Gamma(1-im-i\frac{2\omega}{\tau_H})}
{\Gamma(\half-\beta-im)\,\Gamma(\half-\beta-i\frac{2\omega}{\tau_H})} \,
\tau_H^{\half+\beta-\frac{i}{2}(m+\frac{2\omega}{\tau_H})} \, .
\ea
Note that the normalization is chosen to ensure that the wavefunction equals $1$ at the boundary.
Thus, asymptotically the propagator reduces to
\bea
K(\phi,t,r;\phi',t')&=&\int dm \, d\omega \left(r^{-\half+\beta}
+
\frac{{{\cal B}}(m,\omega)}{{{\cal A}}(m,\omega)} \, r^{-\half-\beta}
\right) e^{-im(\phi-\phi')+i\omega(t-t')}+\ldots \nn \\
&\approx&r^{-\half+\beta}\,\delta(\phi-\phi')\, \delta(t-t')+r^{-\half-\beta}\int dm \, d\omega\
\frac{{{\cal B}}(m,\omega)}{{{\cal A}}(m,\omega)} \, e^{-im(\phi-\phi')+i\omega(t-t')}+\ldots \nn \, .
\ea
The leading behavior
\beq
K(\phi,t,r;\phi',t') \rightarrow r^{-\half+\beta}\,\delta(\phi-\phi')\,\delta(t-t')
\eq
is precisely what we expect from (\ref{Kscaling}), in accordance with the scaling properties of
the scalar field.
Inserting the bulk-to-boundary propagator into
\bea
\label{two-point}
\langle{\cal O}(t_1,\phi_1){\cal O} (t_2,\phi_2)\rangle
&\sim&  \left. \int\ d\phi \, dt \, \sqrt{-g}g^{rr} \,\bar{K}(r,t,\phi;t_1,\phi_1)\,
\partial_r K(r,t,\phi;t_2\phi_2)\; \right|_{r=r_{\text{cutoff}}} \, ,
\ea
with $r_{\text{cutoff}}$ signaling the place where the near-NHEK geometry breaks down,
one can extract the two-point function behavior appropriate for the retarded Green's function:
\bea
\label{two-point-b}
\langle{\cal O}(t_1,\phi_1){\cal O} (t_2,\phi_2)\rangle &\sim&
\left(-\frac{1}{2}+\beta\right) r^{2\beta}\delta(\phi_1-\phi_2)\delta(t_1-t_2) \nn \\
&& + \beta \; \int  dm\ d\omega \; \frac{{{\cal B}}(m,\omega)}{{{\cal A}}(m,\omega)}
e^{-im(\phi_2-\phi_1)+i\omega(t_2-t_1)} + \ldots
\ea
We recognize the term in the first line as a contact term, while the second line is
a contribution to the momentum-space two-point function.
Thus, the retarded Green's function can be read-off directly from (\ref{two-point-b}), and is
\begin{equation}
G_R \sim \frac{{{\cal B}}(m,\omega)}{{{\cal A}}(m,\omega)} \, ,
\end{equation}
in agreement with the prescription (\ref{GRet}).
This concludes our check that the bulk-to-boundary propagator behaves as it should, and reproduces
the results of \cite{Bredberg:2009pv}, as well as \cite{Chen:2010ni}, using the real-time
prescriptions \cite{Son:2002sd,Iqbal:2009fd} encoded by (\ref{GRet}).
Having the right bulk-boundary propagator, we can now
compute the three-point correlator.

\section{Three-point Correlators}

We now come to the main section of our paper, where we compute three-point functions of a scalar
in the near-NHEK region of the Kerr black hole, with the goal of testing the proposed Kerr/CFT correspondence.
Since near-NHEK contains right-moving excitations above extremality, on the conformal field theory side we will
look at finite temperature three-point correlators assuming both a right-moving and a left-moving sector.
We start by considering the gravitational side of the correspondence,
where we adopt the Euclidean prescription (\ref{3ptprescription}) and
make use of our bulk-to-boundary propagator (\ref{bulktoboundary}).
We then move on to the conformal field theory side, where the computation of the
correlation function is straightforward.

In the bulk, we {\it assume} that the scalar field $\Phi$ has a cubic interaction of the schematic form
\beq
\mathcal{G} \int \Phi_{h_1} \Phi_{h_2} \Phi_{h_3} \, ,
\eq
where $h$ is the conformal dimension of the operator ${\cal O}_{h}$ dual to $\Phi$, and
$\mathcal{G}$ denotes the coupling strength.
For simplicity, we focus on operators whose conformal dimensions obey $h_3=h_1+h_2$,
\emph{i.e.} the so-called \emph{extremal} limit which played a prominent role in the
AdS/CFT three-point function literature. This case is particularly simple because the three-point function
reduces to the product of two two-point functions.

We emphasize that, although we are not resorting to the Schwinger-Keldysh formalism of $n$-point correlation
functions \cite{Son:2002sd,Herzog:2002pc,Skenderis:2008dg} but rather using (\ref{3ptprescription}) (with appropriate boundary conditions),
we are able to show agreement between the CFT and gravity side.
As we will see, some interesting features will emerge from the form of the correlation functions, as well
as interesting information on the coupling $\mathcal{G}$ of the interaction.

\subsection{Gravity-side three-point correlation function}

Since we are only working for Matsubara frequencies, in this section we analytically continue
$m$ and $\omega$ -- this won't affect the results of our computations.
Adopting the three-point function prescription (\ref{3ptprescription})
we take as a starting point
\beq
\langle O(t_1,\phi_1)\, O(t_2,\phi_2)\, O(t_3,\phi_3)\rangle
\sim \int d\phi^\prime \, dt^\prime \, dr \,  K_1(r,t^\prime,\phi^\prime;t_1,\phi_1)
K_2(r,t^\prime,\phi^\prime;t_2,\phi_2) K_3(r,t^\prime,\phi^\prime;t_3,\phi_3)\, .
\eq
Using the expression for our bulk-to-boundary propagator (\ref{bulktoboundary}),
the three-point vertex $V_3$ takes the form
\bea
V_3 &\equiv& \langle O(t_1,\phi_1)\, O(t_2,\phi_2)\, O(t_3,\phi_3)\rangle
\nonumber \\
&=& \prod_{i=1}^3 \Bigr(\int dm_i \,e^{-i m_i \phi_i} \, \int d \omega_i \, e^{i \omega_i t_i}\Bigl)
\int_0^{r_c} dr \; \psi_1 \psi_2 \psi_3
\int_0^{2\pi}  d\phi \; e^{i(m_1 + m_2 + m_3)\phi^\prime}
\int_0^{1/T_R} dt \; e^{-i(\omega_1+ \omega_2 + \omega_3)t^\prime} \nn \\
&=& \prod_{i=1}^3 \Bigr(\int dm_i \,e^{-i m_i \phi^\prime_i} \; \int d \omega_i \, e^{i \omega_i t_i^\prime} \Bigl)
\int_0^{r_c} dr \; \psi_1 \psi_2 \psi_3 \; \delta(m_1 + m_2 + m_3) \, \delta(\omega_1+ \omega_2 + \omega_3) \, ,
\ea
where $r_c=r_{cutoff}$ discussed in Section 2.
Note that this is the Fourier transform of the momentum space 3-pt function, $V_3^{m.s.}$.
Instead of performing the integral over the $m_i$ and $\omega_i$ we choose to work in momentum
space, where the three-point function is then given by the following radial integral:
\bea
\label{V3ms}
V_3^{m.s.} &=& \langle O(m_1,\omega_1)\, O(m_2,\omega_2)\, O(m_3,\omega_3)\rangle \nn \\
&=& \delta(m_1 + m_2 + m_3) \, \delta(\omega_1+ \omega_2 + \omega_3)
\int_0^{r_c} dr \; \psi_1 \psi_2 \psi_3 \nn \\
&=& \delta(m_T) \, \delta(\omega_T) N_1 N_2 N_3 \int dr  \,
r^{\frac{m_T}{2}-\frac{\omega_T}{\tau_H}}
\Bigl(1+\frac{r}{\tau_H}\Bigr)^{\frac{m_T}{2}+\frac{\omega_T}{\tau_H}} \times \nn \\
&&\times \prod_{i=1}^3 F\Bigl(\half+\beta_i+m_i,\half-\beta_i+m_i,1+m_i-\frac{2 \omega_i}{\tau_H},-\frac{r}{\tau_H}\Bigr)\, ,
\ea
where $m_T=m_1+m_2+m_3$ and similarly $\omega_T=\omega_1+\omega_2+\omega_3$.
The 3-point function integral (\ref{V3ms}) is difficult to compute exactly.
However, by working in the regime of very small Hawking temperature $\tau_H \ll 1$
and considering the contribution to the integral coming from different regions,
we will be able to show that it reproduces the CFT result.

As will become apparent, it turns out to be convenient to extract the dependence on $\tau_H$ from
the coefficients ${\cal A}, {\cal B}$ defined in (\ref{constants}). We introduce two new constants $A$ and $B$
through
\footnote{Note that our ${\cal A}$ and ${\cal B}$ are, respectively,
$A$ and $B$ in the notation of \cite{Bredberg:2009pv}.}
\beq
\label{AB}
{\cal A} = A\, \tau_H^{\half-\beta-\frac{1}{2}(m+\frac{2\omega}{\tau_H})} \, , \quad \quad
{\cal B} = B \, \tau_H^{\half+\beta-\frac{1}{2}(m+\frac{2\omega}{\tau_H})} \, .
\eq
We break the integration range into the following regions:
\begin{enumerate}
\item \underline{Region I} corresponds to $\frac{r}{\tau_H} \ll 1$.
Here the hypergeometric function can be approximated simply by 1, and the radial wavefunction becomes
\beq
\psi_I \sim N r^{\frac{m}{2}-\frac{\omega}{\tau_H}}
\sim \frac{1}{A} \, \tau_H^{{h-1-\frac{m}{2}+\frac{\omega}{\tau_H}}} \,
r^{\frac{m}{2}-\frac{\omega}{\tau_H}} \, .
\eq
\item \underline{Region II} takes into account the contribution to $V_3$ coming from $r\sim \tau_H$. More precisely, $1-\epsilon < \frac{r}{\tau_H} < 1+\epsilon$,
with $\epsilon < \tau_H$, and we Taylor expand the wavefunction about $z\equiv \frac{r}{\tau_H}=1$:
\beq
\label{Taylor}
\psi_{II} \sim \left. \psi \right\vert_{z=1} + (z-1) \frac{\p \psi}{\p z}|_{z=1}+
\frac{1}{2}(z-1)^2 \frac{\p^2 \psi}{\p z^2}|_{z=1}+\ldots \; .
\eq
\item
\underline{Region III} is $\frac{r}{\tau_H} \gg 1$. Since the argument of the hypergeometric
function is very large (but within the regime of validity
of near-NHEK, $r<r_{\text{cutoff}}$) in this region one can take
\beq
\label{wavefnIII}
\psi_{III}\sim r^{h-1} + \frac{B}{A} \, \tau_H^{2h-1} \, r^{-h} \, ,
\eq
plus corrections that will be \emph{subleading} in the small temperature limit.
As we will see the leading contribution to the 3-point function will come from this region.
\end{enumerate}
%
%
We can now compute the contributions of each region to the momentum-space three-point function vertex,
$V_3^{m.s.} = V_{I} + V_{II} + V_{III}$:
\begin{enumerate}
\item
{\bf Region I:}\\
In this region $r$ is small compared to $\tau_H$, and the 3-point vertex is given by
\bea
V_I &=& \delta(m_T) \; \delta(\omega_T) \int_0^{r_0} dr \; \psi_1 \psi_2 \psi_3 \nn \\
&=& \delta(m_T) \; \delta(\omega_T) \; \frac{1}{A_1 A_2 A_3} \;
\tau_H^{h_1+h_2+h_3-3-\frac{m_T}{2}+\frac{\omega_T}{\tau_H}}
\int_0^{r_0} dr \; r^{\frac{m_T}{2}-\frac{\omega_T}{\tau_H}}  \nn \\
&=& \delta(m_T) \; \delta(\omega_T) \; \frac{1}{A_1 A_2 A_3} \;
\tau_H^{h_1+h_2+h_3-3} \, r_0 \, ,
\ea
where $r_0\ll \tau_H$ and in the last step we made use of the delta functions.\\
\noindent
For the \underline{extremal} case $h_3=h_1+h_2$ this expressions reduces to
\beq
V_I^{extr} = \delta(m_T) \; \delta(\omega_T) \; \frac{1}{A_1 A_2 A_3} \;
\tau_H^{2h_1+2h_2-2} \, \Bigl(\frac{r_0}{\tau_H}\Bigr) \, ,
\eq
which is clearly a finite contribution to the vertex (and is small, since $r_0/\tau_H \ll 1)$.
\item
{\bf Region II:}\\
Here we want to compute
\beq
V_{II} \sim \int_{(1-\epsilon)\tau_H}^{(1+\epsilon)\tau_H} dr \; \psi_1 \psi_2 \psi_3 \, .
\eq
To estimate this term it's enough to Taylor expand $\psi$ about $\frac{r}{\tau_H}=1$.
Looking back at eq. (\ref{Taylor}), we see that we need
$\psi|_{z=1}$ as well as the value of the derivatives there.
To extract the behavior of the wavefunction at 1, it turns out to be convenient to write it in the following way
\footnote{We made use of $_2F_1(a,b;c;x)=(1-x)^{-a} \, _2F_1\left(a,c-b;c;\frac{x}{x-1}\right)$.}:
\beq
\psi(z)=N \, \tau^{-\frac{m}{2}+\frac{\omega}{\tau_H}} \; z^{\frac{m}{2}-\frac{\omega}{\tau_H}}
(1+z)^{\frac{m}{2}+\frac{\omega}{\tau_H}-a} \;\;_2F_1\left(a,c-b;c;\frac{z}{z+1}\right)\, ,
\eq
where $z=\frac{r}{\tau_H}$, $a=\half + \beta+m$, $b=\half-\beta+m$ and $c=1+m -\frac{2\omega}{\tau_H}$.
It is now easy to read off:
\beq
\psi(z)|_{z=1}=N \;\tau_H^{-\frac{m}{2}+\frac{\omega}{\tau_H}} \;\; 2^{\frac{m}{2}+\frac{\omega}{\tau_H}-a}
\;\;_2F_1\left(a,c-b;c;\frac{1}{2}\right)\, .
\eq
The crucial point to notice is that
\beq
\left. \psi(z)\right |_{z=1} \propto \frac{1}{A} \;\; \tau_H^{h-1-m+\frac{2\omega}{\tau_H}}\, ,
\eq
where the proportionality constant doesn't affect the way this term scales with the temperature $\tau_H$.
Similarly, we can show that the derivatives of $\psi$ scale in the same way with temperature,
\beq
\left.\frac{\p \psi}{\p z}\right |_{z=1} \propto \frac{1}{A} \; \tau_H^{h-1-m+\frac{2\omega}{\tau_H}}\, , \;\;\;\;
\left.\frac{\p^2 \psi}{\p z^2}\right |_{z=1} \propto \frac{1}{A} \; \tau_H^{h-1-m+\frac{2\omega}{\tau_H}}\, .
\eq
Thus, the Taylor expansion of the radial wavefunction can be written schematically as
\beq
\psi_{II} = \frac{1}{A} \; \tau_H^{h-1-m+\frac{2\omega}{\tau_H}}
\bigl[C_1 + (z-1)C_2 + (z-1)^2 C_3 \bigr]
\eq
where $C_1,C_2,C_3$ are constants that don't play any role in the scaling with $\tau_H$.
The 3-point vertex from region II then becomes
\bea
V_{II} &=&\delta(m_T) \; \delta(\omega_T) \; \tau_H \int_{1-\epsilon}^{1+\epsilon} dz \; \psi_1 \psi_2 \psi_3 \nn \\
&=&  \frac{1}{A_1 A_2 A_3} \; \delta(m_T) \; \delta(\omega_T)
\; \tau_H^{h_1+h_2+h_3-2} \nn \\
&& \times
\int_{1-\epsilon}^{1+\epsilon} dz
\Bigl[\mbox{constant} + (z-1) \, \mbox{constant} + (z-1)^2 \, \mbox{constant} + \ldots \Bigr] \nn \\
&=& \frac{1}{A_1 A_2 A_3} \; \delta(m_T) \; \delta(\omega_T)\; \tau_H^{h_1+h_2+h_3-2}
\Bigl[\epsilon \times \mbox{constant} + {\cal O} (\epsilon^3)\Bigr]\,.
\ea
Thus, for \underline{extremal} conformal weights we find that the contribution to the vertex is
\beq
V_{II} = \frac{1}{A_1 A_2 A_3} \; \delta(m_T) \; \delta(\omega_T)\; \tau_H^{2 h_1+2 h_2-2}
\Bigl[\epsilon \times \mbox{constant} + {\cal O} (\epsilon^3)\Bigr]\,,
\eq
which is again finite (and small since $\epsilon \ll 1$).
\item
{\bf Region III}\\
When $r\gg \tau_H$, the wavefunction can be approximated by (\ref{wavefnIII}), and the 3-point vertex
gives the following terms:
\bea
\label{V3generalweights}
V_{III} &=& \int_{r_{1}}^{r_{\text{cutoff}}} dr \; \psi_1 \psi_2 \psi_3 \nn \\
&=& \Biggl\{ \frac{B_1 B_2 B_3}{A_1 A_2 A_3} \; \frac{r^{-h_1-h_2-h_3+1}}{1-h_1-h_2-h_3}\ \tau_H^{2h_1+2h_2+2h_3-3}\ +\  \frac{r^{h_1+h_2+h_3-3}}{h_1+h_2+h_3-3} \nn \\
&&+ \Biggl[ \frac{B_1 B_2}{A_1 A_2} \; \frac{r^{h_3-h_1-h_2}}{h_3-h_1-h_2}\ \tau_H^{2h_1+2h_2-2} \;
+ \frac{B_2 B_3}{A_2 A_3} \;  \frac{r^{h_1-h_2-h_3}}{h_1-h_2-h_3} \ \tau_H^{2h_2+2h_3-2} \;\nn \\
&&\qquad\qquad  \qquad  \qquad  \qquad  \qquad \qquad  \quad +\frac{B_1 B_3}{A_1 A_3}  \; \frac{r^{h_2-h_1-h_3}}{h_2-h_1-h_3}\; \tau_H^{2h_1+2h_3-2}\Biggr] \nn \\
&&+\Biggl[ \frac{B_1}{A_1}  \; \frac{r^{h_2+h_3-h_1-2}}{h_2+h_3-h_1-2}\; \tau_H^{2h_1-1}
+ \frac{B_2}{A_2}  \; \frac{r^{h_1+h_3-h_2-2}}{h_1+h_3-h_2-2} \; \tau_H^{2h_2-1}\nn \\
&&\ \  \qquad \qquad  \qquad  \qquad  \qquad  \qquad \qquad +
\frac{B_3}{A_3} \; \frac{r^{h_1+h_2-h_3-2}}{h_1+h_2-h_3-2}\; \tau_H^{2h_3-1} \Biggr] \Biggr\}
\Bigg |_{r_{1}}^{r_{\text{cutoff}}}\, , \nn\\
\ea
where we take the lower bound $r_{1}$ to be sufficiently bigger than $\tau_H$.
The cutoff $r_{\text{cutoff}}$ denotes the distance at which the geometry is no longer near-extremal.
We should mention that we expect to be able to take $r_{\text{cutoff}}\rightarrow \infty$ for the near-horizon region of extremal rotating black holes.
We can immediately make some comments about the structure of these terms.
First of all, since we expect
each conformal weight to be
\footnote{This follows from the fact that we know that the A piece of
the wavefunction (\ref{wavefnIII}) is non-normalizable at the boundary, requiring $h>1$.}
$h_i\geq 1$,
all the terms containing a ratio of the form $BB/AA$, as well as the single $BBB/AAA$ term,
are finite at the boundary (provided $h_i+h_j>h_k$ for all pairs of weights).
The remaining terms (of the form $B/A$ or without any $A$ or $B$) correspond to
contact terms, as can be shown explicitly by
going back to coordinate space
\footnote{For example, note that since the $B_1/A_1$ term contains only dependence on $(m_1,\omega_1)$, its
Fourier transform over $(m_2,m_3,\omega_2,\omega_3)$ is trivial, and
yields a delta function of the form $\delta(t_2-t_3) \delta(\phi_2-\phi_3)$.}.

\indent
We now restrict our attention to the \underline{extremal} case $h_3=h_1+h_2$.
Neglecting all contact terms, the vertex from region III can be written in a suggestive way as:
\bea
\label{VIII}
V_{III} &\sim& \tau_H^{2h_1+2h_2-2} \Biggl[
\frac{1}{1-2h_1-2h_2} \; \frac{B_1 B_2 B_3}{A_1 A_2 A_3} \; \Bigl(\frac{\tau_H}{r_1}\Bigr)^{2h_1+2h_2-1}  \nn \\
&+&\frac{1}{h_3-h_1-h_2} \; \frac{B_1 B_2}{A_1 A_2} \; \Bigl(\frac{\tau_H}{r_1}\Bigr)^{0}
-\frac{1}{2h_1} \; \frac{B_1 B_3}{A_1 A_3} \; \Bigl(\frac{\tau_H}{r_1}\Bigr)^{2h_1}
-\frac{1}{2h_2} \; \frac{B_2 B_3}{A_2 A_3} \; \Bigl(\frac{\tau_H}{r_1}\Bigr)^{2h_2} + \ldots
\Biggr]\, . \nn \\
\ea
The terms we dropped are positive powers of $\frac{\tau_H}{r_{\text{cutoff}}}$ and are therefore significantly
smaller (and vanish if the cutoff were taken to infinity).

However, there is an important subtlety involved in dealing with the $\frac{B_1B_2}{A_1A_2}$ term.
If one imposed the extremality condition before evaluating the integral, the result would be
a logarithmic term.
However, a recipe for computing extremal correlators which has emerged from the AdS/CFT literature suggests
that the ``extremal limit'' $h_3 \rightarrow h_1+h_2$ should be taken only \emph{after} evaluating the integral.
In the AdS/CFT literature, this is the statement that
\bea
\langle {\cal O}_{h_2+h_3}(x_1) {\cal O}_{h_2}(x_2){\cal O}_{h_3}(x_3)\rangle
&=& \frac{C_{h_2+h_3,h_2,h_3}}{|x_1-x_2|^{2h_2}|x_1-x_2|^{2h_3}} \, \nn \\
\text{where} \;\;C_{h_2+h_3,h_2,h_3} &=& \text{Lim}_{h_1 \rightarrow h_2+h_3}
\; C_{h_1,h_2,h_3}\, ,
\ea
\emph{i.e.} the answer for extremal correlators should be read off from the $h_3 \rightarrow h_1+h_2$ limit
of the non-extremal one. This is the procedure we follow in evaluating (\ref{V3generalweights}),
which can be thought of as an \emph{analytic continuation} from non-extremal correlators.
We find that the $\frac{B_1B_2}{A_1A_2}$ term contains a factor $\sim \frac{1}{h_3-h_1-h_2}$
that diverges precisely in the extremal limit, in analogy
with previous AdS/CFT computations.
This result seems to put a constraint on the
overall coupling of the interaction
\beq
\mathcal{G} \sim h_3-h_2-h_1,
\eq
which will ensure that the complete result for the three-point function is finite and in agreement with conformal field theory.
Finally, note that the remaining terms in (\ref{VIII}) are all finite
and small, since by assumption $r/\tau_H \gg 1$ and all the conformal weights are positive.
\end{enumerate}
In summary, adding together the contributions to the vertex coming from different regions, we find
that the gravity-side three-point correlator in momentum space is
\bea
\label{Vfinal}
V_3^{m.s.} &\sim& \tau_H^{2h_1+2h_2-2}\Biggl[ \biggl(\frac{1}{h_3-h_1-h_2}\biggr) \frac{B_1B_2}{A_1A_2} +
\mbox{finite terms} \Biggr] \delta(m_T) \, \delta(\omega_T) \nn \\
&=& \Biggl[\biggl(\frac{1}{h_3-h_1-h_2}\biggr)
\frac{{\cal B}(m_1,\omega_1) \,{\cal B}(m_2,\omega_2)}{{\cal A}(m_1,\omega_1)\,{\cal A}(m_2,\omega_2)}
+ \ldots \Biggl]\delta(m_T) \, \delta(\omega_T)
\ea
in the limit $h_3 \rightarrow h_1+h_2$.

We finish this section with a remark. In the context of AdS/CFT it was possible to {\it calculate} precisely a behavior such as the one mentioned above for the coupling constant
$\mathcal{G}$, where integration over the spherical harmonics of $S^5$  gave such a result
\cite{Freedman:1998tz}.
Kerr/CFT is, on the other hand, a three-dimensional duality that does not necessarily involve string theory.
Thus the information on the coupling can only be obtained by matching with the conformal field theory.
It is very plausible that string theory would be able to explain the value of these couplings at a
more fundamental level, if an embedding of Kerr/CFT into string theory were found.

\subsection{Finite temperature CFT three-point correlation function}

According to the Kerr/CFT duality the near-NHEK background should be dual to a two-dimensional conformal
field theory with temperature $T_L=1/2\pi$ for the left-moving sector and $T_R=\frac{\tau_H}{4\pi\lambda}$
for the right-movers. We recall from Section \ref{KerrCFTSection} that $T_L$ is
the Frolov-Thorne temperature whereas the right-movers feel a finite $T_R$ only in the near-horizon
black hole region.
Thus, to compare our gravitational result (\ref{Vfinal}) to the conformal field theory side, we need
the contribution of both right- and left-moving sectors to finite-temperature three-point correlators.

Finite temperature correlators can be constructed from their zero-temperature counterparts
by making the corresponding coordinates periodic, with the circumferences equal to inverse temperatures.
A three-point correlator for a primary chiral field at zero temperature is given by
\bea
\langle {\cal O}_1(x^+_1, x^-_1){\cal O}_2(x^+_2, x^-_2){\cal O}_3(x^+_3, x^-_3)\rangle&=&C_{123}
\frac{1}{(x^+_{12})^{h_1+h_2-h_3}(x^+_{23})^{h_2+h_3-h_1}(x^+_{13})^{h_3+h_1-h_2}}\nn\\
&&\times\frac{1}{(x^-_{12})^{\bar h_1+\bar h_2-\bar h_3}(x^-_{23})^{\bar h_2+\bar h_3-\bar h_1}
(x^-_{13})^{\bar h_3+\bar h_1-\bar h_2}}\ ,
\ea
where $h_i,\bar h_i$ are the conformal dimensions of ${\cal O}_i$, and $x_{ij}\equiv x_i-x_j$.
One can map the infinite plane
to a torus with coordinates $(t^-,t^+)$ via
\begin{equation}
x^-=e^{2\pi i T_L t^- } \ , \qquad  x^+=e^{2\pi i  T_R t^+} \ .
\end{equation}
Using the coordinate transformation behavior of primary fields
\begin{equation}
{\cal O}'(t^-,t^+)=\left(\frac{dt^-}{d x^-}\right)^{-h}\left(\frac{dt^+}{d x^+}\right)^{-\bar h}{\cal O}(x^-,x^+)
\end{equation}
one finds
\footnote{Notice that we are not explicitly including the chemical potential, because one can absorb it
by appropriately shifting the frequencies.}
\begin{eqnarray}
\label{cftcoord}
\langle{\cal O}_1(t^-_1, t^+_1)&&{\cal O}_2(t^-_2, t^+_2)\  \ {\cal O}_3(t^-_3, t^+_3)\rangle=C_{123}
\times \nn \\
&&\left( \frac{\pi T_L}{\sin(\pi T_Lt^-_{12})}\right)^{h_1+h_2-h_3}
\left( \frac{\pi T_L}{\sin(\pi T_Lt^-_{23})}\right)^{h_2+h_3-h_1}\left( \frac{\pi T_L}{\sin(\pi T_Lt^-_{13})}
\right)^{h_1+h_3-h_2}\nn\\ &&
\left( \frac{\pi T_L}{\sin(\pi T_Rt^+_{12})}\right)^{\bar h_1+\bar h_2-\bar h_3}
\left( \frac{\pi T_R}{\sin(\pi T_Rt^+_{23})}\right)^{\bar h_2+\bar h_3-\bar h_1}
\left( \frac{\pi T_R}{\sin(\pi T_Rt^+_{23})}\right)^{\bar h_1+\bar h_3-\bar h_2}
\end{eqnarray}
Recall that on the bulk side we are are computing the three-point function in momentum space.
We therefore Fourier transform (\ref{cftcoord}) and obtain \emph{for each sector}
\bea
\langle{\cal O}(\omega_{1}){\cal O}(\omega_{2}){\cal O}(\omega_{3})\rangle=\int dt_1 dt_2 dt_3 \ e^{i \omega_1 t_1+i \omega_2 t_2+i \omega_3 t_3}
\langle{\cal O}(t_{1}){\cal O}(t_{2}){\cal O}(t_{3})\rangle \ .
\ea
In the extremal case $h_i=h_j+h_k$ the three-point correlator (\ref{cftcoord}) simplifies significantly.
Without loss of generality we choose the case $h_3=h_1+h_2$:
\bea\label{cftres}
&&\langle{\cal O}(\omega_{1}){\cal O}(\omega_{2}){\cal O}(\omega_{3})\rangle=\int dt_1 dt_2 dt_3
\ e^{i \omega_1 t_1+i \omega_2 t_2+i \omega_3 t_3}\left(\frac{\pi T_L}{\sin(\pi T_L t_{23})}\right)^{2h_2}
\left(\frac{\pi T_L}{\sin(\pi T_L t_{13})}\right)^{2h_1}\nn\\
&&=2 \, \pi \, \delta(\omega_1+\omega_2+\omega_3)\int_0^{1/{T_L}}  dt_{23}  \ e^{i\omega_2 t_{23}}
\left(\frac{\pi T_L}{\sin(\pi {T_L} t_{23})}\right)^{2h_2}\int_0^{1/{T_L}} dt_{13}
\ e^{i\omega_1 t_{13}}\left(\frac{\pi {T_L}}{\sin(\pi {T_L} t_{13})}\right)^{2h1} \, ,\nn \\
\ea
with an analogous expression for the right-moving sector, with
the corresponding right-moving temperature $T_R$.
As in (\ref{EuclG}), the frequencies are forced to be integers
\begin{equation}
\omega=2\pi k T \ ,  \nn \end{equation}
with $k$ an integer.
Notice that the expression (\ref{cftres}) is equal to a product of two 2-point correlators, which is again
a consequence of having chosen extremal conformal weights.
Finally, the left- and right-moving sectors combined together yield
\bea
\langle{\cal O}(\omega_{L1},\omega_{R1}){\cal O}(\omega_{L2},\omega_{R2}){\cal O}(\omega_{L3},\omega_{R3})
\rangle&=& (2\pi)^2\delta(\omega_{L1}+\omega_{L2}+\omega_{L3})\delta(\omega_{R1}+\omega_{R2}+\omega_{R3}) \nn \\
&&\times \langle{\cal O}(\omega_{L1},\omega_{R1}){\cal O}(0,0)\rangle \ \
\langle{\cal O}(\omega_{L2},\omega_{R2}){\cal O}(0,0)\rangle \, .
\nn \\
\ea
The final step required to relate the CFT result with the one coming from the bulk
is the identification of the gravity-side parameters -- the angular momentum and frequency $m,\omega$ --
with the CFT frequencies $\omega_L, \omega_R$.
However, this was already done by \cite{Bredberg:2009pv}, who identified
\beq
\omega_L=\omega,\quad \omega_R=m \, ,
\eq
in order to match the two-point correlators.
In addition, here we have found that the coupling must have the specific form
\beq
\label{coupling}
\mathcal{G} \propto h_3-h_1-h_2 \, ,
\eq
which again ensures agreement between the bulk and boundary calculations.

To summarize, we can write the final expression for the CFT three-point correlator in a more compact and elegant way by
recalling that each two-point function is simply given by ${\cal B}/{\cal A}$:
\begin{equation}
\label{CFTFinalVertex}
\langle{\cal O}(\omega_1,m_1){\cal O}(\omega_2,m_2){\cal O}(\omega_3,m_3)\rangle\sim \delta(m_1+m_2+m_3)
\delta(\omega_1+\omega_2+\omega_3)\frac{{\cal B}(m_1,\omega_1)}{{\cal A}(m_1,\omega_1)}
\frac{{\cal B}(m_2,\omega_2)}{{\cal A}(m_2,\omega_2)},
\end{equation}
which is precisely the same result we found in the bulk calculation of the previous section!

\section{Discussion}
\label{discussionsection}

Our main interest in this note was to provide further tests of the conjectured Kerr/CFT correspondence
for the case when there is some excitation energy above extremality, i.e. when both right and left moving
sectors are excited.
In particular, we wanted to ask whether it would be possible to extend the results of \cite{Bredberg:2009pv}
-- who were able to match two-point correlation functions -- to three-point correlators.
Furthermore, we were interested in whether one could naively use the \emph{Euclidean} prescription for computing
three-point functions in the near-NHEK background,
and find agreement with finite-temperature CFT correlation functions.
The answer to the latter question is yes, at least for the special class of operators we restricted our analysis
to. The more general case of arbitrary conformal dimensions is work in progress.

From our final expressions for the three-point correlators -- (\ref{Vfinal}) for the gravitational three-point
function and (\ref{CFTFinalVertex}) for the CFT side -- we see that the result
is identical, provided a specific behavior of the coupling is assumed.
The gravitational result
\beq
V_3 \sim \biggl(\frac{1}{h_3-h_1-h_2}\biggr)
\frac{{\cal B}(m_1,\omega_1) \,{\cal B}(m_2,\omega_2)}{{\cal A}(m_1,\omega_1)\,{\cal A}(m_2,\omega_2)}
\eq
contains an additional factor which seemingly diverges in the $h_3 \rightarrow h_1+h_2$ limit.
Interestingly, this factor is reminiscent of three-point function calculations in
$AdS_5 \times S^5$ \footnote{As well as in its $AdS_3\times S^3 (\;\times \;T^4 \;\mbox{or}\; K3)$ counterpart.},
where it was also a feature of extremal correlators $h_3=h_1+h_2$.

In the $AdS_5 \times S^5$ context, after dimensional reduction of the field equations of Type IIB SUGRA,
scalar fields (chiral primaries) $s_h$ were identified \cite{Lee:1998bxa}
which have an effective
cubic interaction of the form
$${\cal G}(h_1,h_2,h_3) \, \int_{AdS_5}  s_{h_1} s_{h_2} s_{h_3}\; ,$$
with the $h_i$ denoting the conformal weights of the dual operators.
For extremal correlators with $h_3 \rightarrow h_1+h_2$ the $AdS_5$ vertex
$\int_{AdS_5} s_{h_1} s_{h_2} s_{h_3}$ was shown to diverge as $\frac{1}{h_3-h_1-h_2}$ \cite{Freedman:1998tz}.
However, the bulk coupling ${\cal G}(h_1,h_2,h_3)$ contains a prefactor
$\propto h_3-h_1-h_2$, which vanishes when $h_3=h_1+h_2$. This in turn balances the divergence of
the bulk integral, yielding a finite answer for the correlator --
the subtleties of this extremal limit were discussed in \cite{D'Hoker:1999ea}.
The presence of this ``vanishing prefactor'' is a well-known feature of AdS/CFT three-point correlators,
and it is intriguing that the same type of structure appears
in the Kerr/CFT context.

Unlike the $AdS_5 \times S^5$ case, where the coupling of the chiral primaries could be determined by
dimensional reduction and is therefore related to UV physics, the theory describing the near-NHEK geometry
is a three-dimensional theory which does not necessarily involve string theory.
If a string theory embedding of Kerr/CFT was known, it would provide us with a geometrical way to
obtain information about the coupling. However, at the moment such a construction is lacking, and we can't derive
$\mathcal{G}$ from first principles. We leave such an interesting question for future work.
What our result seems to suggest, however, is that in order to have a well-defined finite contribution to
the extremal three-point correlator, the coupling of the cubic interaction should contain
a vanishing prefactor of the form\footnote{We are grateful to K.Skenderis for pointing out to us that the 
vanishing of the extremal coupling is expected from the structure of conformal anomalies, in theories that admit a Coulomb branch
(see e.g. \cite{Skenderis:2006uy,Petkou:1999fv}). We refer the reader to \cite{Becker:2010dm} for a more detailed discussion in the Kerr/CFT context.}
\beq
\label{coupling}
\mathcal{G} \propto h_3-h_1-h_2 \, ,
\eq
which would compensate the divergence of the bulk integral, in analogy
with the standard AdS/CFT literature.
In the latter case such a form of the coupling could be specifically calculated and
obtained from compactification (in the $AdS_5 \times S^5$ case from the integration over the $S^5$).
We should emphasize that the divergence of the bulk integral occurs already in pure AdS space,
and is therefore not strictly related to the near-NHEK geometry.
One can try to better understand this issue -- and its implications for the structure of the coupling --
in the simpler AdS context, where regularization schemes
along the lines of holographic renormalization \cite{de Haro:2000xn}
might prove useful.
A calculation in which the extremal value of the coupling is taken from the very beginning  --
in analogy with the work of {\cite{D'Hoker:1999ea}} in AdS/CFT -- might be an interesting question for
future research.

The fact that we found agreement of extremal correlation functions is remarkable for several reasons.
First of all, we were able to obtain agreement by adopting the simple Euclidean recipe (\ref{3ptprescription})
and avoiding the various subtleties associated with the Schwinger-Keldysh formalism.
Second, our result seems to indicate that a non-renormalization theorem is at hand.
Such non-renormalization theorems appeared in the context of AdS/CFT in
\cite{Freedman:1998tz}, where a large amount of supersymmetry is available. The Kerr/CFT correspondence is a relation between non-supersymmetric theories (at finite temperature), yet rather surprisingly a non-renormalization theorem seems to be at play.
Thus, although this is only a first step towards the computation of more general three-point functions,
our matching is a step in the direction of extending AdS/CFT to non-supersymmetric settings.

In this note we have restricted our attention to the case of extremal correlators. We are currently
exploring whether our analysis can be extended to the case of generic conformal weights.
Furthermore, our treatment of the three-point function on the gravity side should be easily generalizable
to the near-horizon region of any extremal, rotating black hole. This is work in progress, which
will not be addressed here.
 Finally, would be interesting if one could extract additional information about the conformal field theory
dual to near-NHEK (and the types of operators one is exciting) from knowledge of  couplings of the form
(\ref{coupling}) and its generalizations.

\acknowledgments

We would like to thank Katrin Becker, Geoffrey Compere, Dan Freedman, Gaston Giribet, Guangyu Guo, Tom Hartman, Finn Larsen, Albion Lawrence, Jim Liu,
Andy Strominger, Jan Troost and Anastasia Volovich for many useful conversations.
We are especially grateful to Andy Strominger for comments on the draft.
The work of M.B. and W.S. was supported by NSF under grant PHY-0505757 and the University of Texas A\&M.
The work of S.C. has been supported by the Cambridge-Mitchell Collaboration in Theoretical
Cosmology, and the Mitchell Family Foundation. M.B. and W.S. would like to thank
CECS (Valdivia) and the organizers of the 3D gravity workshop for a very stimulating meeting, where this work was presented. M.B.would further like to A.Lawrence and B.Lian for a great meeting at Brandeis where this work was recently presented.

\end{document}